\documentclass[conference]{IEEEtran}
\IEEEoverridecommandlockouts
\usepackage{cite}
\usepackage{amsmath,amssymb,amsfonts}
\usepackage{algorithmic}
\usepackage{graphicx}
\usepackage{textcomp}
\usepackage{xcolor}
\def\BibTeX{{\rm B\kern-.05em{\sc i\kern-.025em b}\kern-.08em
    T\kern-.1667em\lower.7ex\hbox{E}\kern-.125emX}}
\begin{document}

\title{FlowBazaar: A Market-Mediated Software Defined Communications Ecosystem at the Wireless Edge
}

\author{\IEEEauthorblockN{Rajarshi Bhattacharyya, Bainan Xia, Desik Rengarajan, Srinivas Shakkottai, Dileep Kalathil}
\IEEEauthorblockA{\textit{Department of Electrical and Computer Engineering} \\
\textit{Texas A\&M University}\\
College Station, Texas \\
\{rajarshibh, xiabainan, desik, sshakkot, dileep.kalathil\}@tamu.edu}
}

\maketitle

\begin{abstract}
The predominant use of wireless access networks is for media streaming applications, which are only gaining popularity as ever more devices become available for this purpose.  
However, current access networks treat all packets identically, and lack the agility  to determine which clients are most in need of service at a given time.  
Software reconfigurability of networking devices has seen wide adoption, and this in turn implies that agile control policies can be now instantiated on access networks.
The goal of this work is to design, develop and demonstrate FlowBazaar, an market-based approach to create a value chain from the application on one side, to algorithms operating over reconfigurable infrastructure on the other, so 
that applications are able to obtain necessary resources for optimal performance.
Using YouTube video streaming as an example, we illustrate how FlowBazaar is able to adaptively provide such resources and attain a high QoE for all clients at a wireless access point.
\end{abstract}

\section{Introduction}
\label{intro}

A majority of Internet usage today occurs over wireless access networks, and this trend is only likely to accelerate with the growing penetration of connected televisions, VR headsets, and other smart home appliances.   These access networks are growing ever more dense, and the difference between WiFi and cellular access is becoming less clear as 5G standards that require small, densely located cells, and next generation WiFi standards that utilize per-packet scheduling rather than random access become more popular.  

A major fraction of the packets carried by these wireless access networks are related to media streaming, which have relatively stringent constraints on the required quality of service (QoS) provided by the network for ideal operation.  These QoS metrics typically are measured as link statistics such as $[Throughput,\ RTT,\ Jitter,\ Loss Rate].$  The impact of such QoS on user satisfaction is identified in terms of Quality of Experience (QoE).   QoE is measured as a number in the interval $[1, 5],$ and can be dependent on the application and its evolving state.   For example, the application can be video streaming over the Web, with the state being the number and duration of stalls (re-buffering events) that have been experienced thus far.  Supporting a large number of concurrent streams of this kind, while ensuring high QoE for all clients is a major challenge.

\begin{figure}[htbp]
\begin{center}
\includegraphics[width=3in]{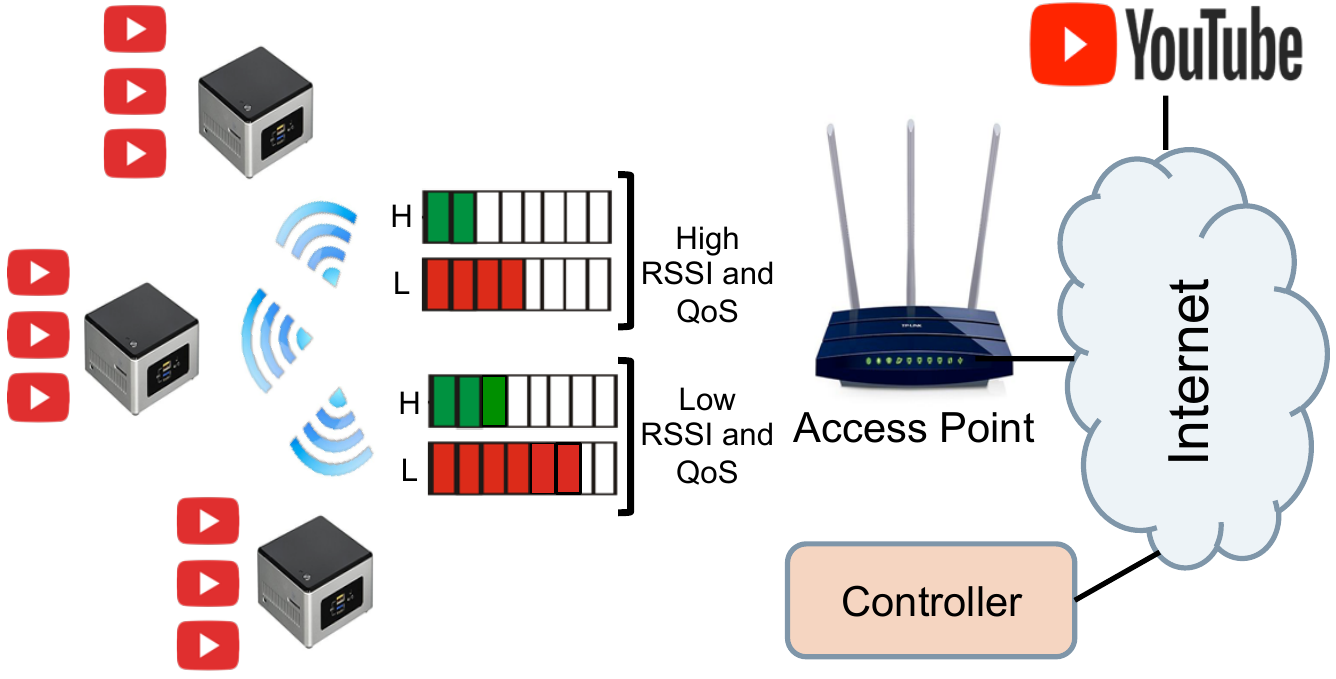}
\caption{Ensuring high QoE video streaming via adaptive prioritization.
}
\label{fig:setup}
\end{center}
\vspace{-0.1in}
\end{figure}
As a concrete example, consider Figure~\ref{fig:setup} that shows nine simultaneous YouTube clients that are supported over a wireless access network.  This setup is used for our laboratory experiments, and can emulate a range of load and channel conditions by restricting the available QoS values at the access point.   The traditional (vanilla) approach is to maintain a single queue, and to treat all packets identically regardless of the importance of the packets to the QoE of the clients.  So a session that has already buffered up many seconds of video might get equal service as one that is near stalling.  While this approach might be acceptable when the number of streams is limited, the need to support multiple high quality streams motivates the desire to do better.

Given that queuing behavior is fundamental to all elements of the QoS statistics mentioned above, differentiated queuing at the access point immediately suggests itself.  Token-bucket-based shaping can be used to create high-priority and low-priority queues, with the QoS statistics of the former being superior to that of the latter.  Furthermore, we can create multiple ``bins'' of queues as shown in Figure~\ref{fig:setup}, with each bin corresponding to similar client channel conditions (with worse channels implying lower achievable QoS), and allocate them similar time-spectrum resources.   Then a basic question is that of periodically deciding client schedules: \emph{ Given the current QoE and video state at each client, how should the controller assign clients to queues for the next decision period?}

If we visualize the system state as being the QoE, stalls and buffered video of all clients, and actions as deciding assignments of clients to queues, then the problem of maximization of the sum QoE can be posed as a Markov Decision Process (MDP).  However, the transition kernel of this MDP is unknown apriori.  In \cite{bhattacharyya2019qflow}, we considered this problem from the perspective of reinforcement learning (RL), and utilized model-based and model-free RL approaches to  finding the optimal policy over a platform entitled QFlow.  Indeed, impressive improvements were seen on QFlow---in a situation where the vanilla policy was only able to achieve a QoE of about 3, the RL approaches achieved a QoE of over $4.7$.

However, there are two main drawbacks to the RL approach.  First, is complexity.  The model-free approach using Q-learning requires an enormous amount of training, which necessitates its being trained over a simulated data set.  Furthermore, the model-based approach needs to obtain the complete transition kernel of the system, which does not scale well as the number of clients increases.  The second is a question of incentives.  The state of each client has to be supplied by the clients themselves, which implies that intelligent clients could obtain more than their fair share of resources through appropriate state declarations to the AP.  

The goal of this paper is to design, develop and validate FlowBazaar, an auction-based system for ensuring high QoE video streaming using information volunteered by the clients themselves.  FlowBazaar uses the same hardware platform as QFlow, but the focus is on eliciting true value functions from the clients, and using that to decide on prioritization policy.  

\subsection*{Main Results and Organization}



In this paper, we design FlowBazaar as a scheme for ensuring high QoE of video streaming.  The system consists of a setup similar to Figure~\ref{fig:setup} under which all TCP flows corresponding to a particular client can be periodically assigned to any one of the queues.  The system of setting up and assigning flows to queues follows \cite{bhattacharyya2019qflow}, in which OpenFlow extensions are developed to enable an OpenFlow controller to instantiate policy decisions at the AP.  The decision period is set as 10 seconds, so as to allow time for the TCP flows to attain equilibrium.  
Each client samples its video state, and maps this state to a standard video QoE model called Delivery Quality Score (DQS) \cite{7025402}.    The QoE forms the one-step reward seen by the client.

The core of FlowBazaar is an incentive compatible (truth-telling) auction that is conducted every 10 seconds.  In our setup, we only permit two clients to be assigned to the high priority queue (in each bin), and hence select a third-price auction as our mechanism for queue assignment.  Here, the clients are asked for bids, and the top two bidders are selected, with the price being charged equal to the third highest value.  It is easy to show that such an auction will elicit the true value functions of the clients.  Bids for the auction are placed via a smart middleware algorithm \emph{(not by a human end-user, who may be unaware of the existence of the system)}.  The bids themselves are values in the interval $[0,5],$ and can be interpreted as the number of cents that the bidding algorithm is willing to pay for high priority service for the next 10 seconds.  We calculate that the eventual dollar price paid will be of the order to a few tens of dollars per month, consistent with cellular data access billing schemes of today.

Under this system, clients first learn the system model in an offline manner, i.e., the marginal transition kerenels that correspond to the change in state at a client,  given the queue that it was assigned to.  Once the kernel is known, the client solves an MDP to determine its bid, given a \emph{belief distribution} of bids made by other clients.   This belief is simply the empirical distribution of bids made thus far (using a weighted moving average to eliminate older information)  and is collected and made available by the controller to all clients.  Our earlier work shows analytically in a similar problem framework that an efficient equilibrium in the space of belief distributions exists under the so called mean field approximation~\cite{ManRam14}.

FlowBazaar, solves the two main issues of the RL approach by leaving the MDP solution to the clients themselves under an auction-based scheduling scheme.  Here, the clients only need to consider the \emph{marginal transition kernel} (impact on their own states of the scheduling decision) while solving the MDP, and coordinate with one another via the auction conducted periodically at the access point.  While one might think that the resulting allocation is suboptimal, it actually turns out to yield a superior QoE over reinforcement learning for all clients.  This result suggest that a indexing of state is occurring in this problem, under which each client state is associated with a  real-number index.  The optimal policy simply picks the clients with largest indices to promote to the high priority queues.  We empirically validate this hypothesis, and find that such an index policy performs as well or better than all others, lending credence to the indexing claim.

The paper is organized as follows. 
Section \ref{section:related} reviews existing studies on QoS and queueing, OpenFlow extensions to wireless and auctions and pricing on the wireless edge.  
Section \ref{section:system} describes our models for the system-wide model for overall QoE maximization, as well as the individual model for the auction.  
Section \ref{architecture} describes the system architecture of FlowBazaar, emphasizing the algorithmic nature of the auctions with no human intervention.  
Section \ref{section:policy}  determines both the system-wide and marginal kernels experimentally, and designs policies that apply to the different cases and belief structures.  
Section \ref{section: experiments} presents our experimental results, that clearly illustrate the superior performance of FlowBazaar over tall other policies that we evaluated against.  
Finally, Section \ref{section:conclusion} concludes the paper.

\section{Related Work}
\label{section:related}


\emph{Optimal Queueing:} There has been significant work on QoS as a function of the scheduling policy, e.g., a sequence of work starting with \cite{TasEph_92}, and follow on work in the wireless context that resulted in algorithms such as backpressure-based scheduling and routing in wireless networks \cite{ErySriPer_05} and more recently \cite {HouBor09} that ensures that strict delay guarantees are met.   Most of these works aim at maximizing throughput or loss rate, but they do not consider all the elements of QoS together. Also, they do not map received QoS to application QoE.

\emph{Auctions and Scheduling:} There has also been work on using price or auction-based resource allocation in the wireless context. On the analytical side, \cite{Auction06} considered the problem of auction-based wireless resource allocation.  Here, users participate in a second price auction and bid for a channel.  It was shown that with finite number of users, a Nash Equilibrium exists and the solution is Pareto optimal.  In \cite{ManRam14}, an auction framework is presented in which queues (representing apps on mobile devices) repeatedly bid for service in a second-price auction that determines which set of queues will be selected for service.  They show that under a large system scaling (called the mean field game regime), the result of the auction would be the same as that of the longest-queue-first algorithm, and hence ensuring fair service for all.  Our design of auction-based scheduling algorithms are motivated by these ideas.  In the context of experiments, a recent trial of a price-based system is described  in \cite{HaSen12}.  Here, day-ahead prices are announced in advance to users, who can choose to use their cellular data connection based the current price.   Thus, the decision makers are the human end-users that essentially have an on/off control.   Furthermore, the prices are not dynamic and have to be determined offline based on historical usage.

\emph{OpenFlow Extensions:} There has been significant research into the development of OpenFlow extension to cross-layer wireless configuration selection.  In this context, CrossFlow \cite{CrossFlow1,CrossFlow2} uses the SDN framework for configuring software defined radios.  Similarly \AE therFlow~\cite{Muxi}, extends OpenFlow for enabling remote configuration of WiFI access points.  Finally, recent systems such as  AeroFlux \cite{aeroflux} and OpenSDWN \cite{schulz2015opensdwn}  enable packet prioritization for flows that are identified by packet inspection as belonging to high priority applications, such as video streaming.  However, these are all offline static policies in that they do not relate the prioritization policy with the state of the application.

\section{System Model}
\label{section:system}

We consider a resource constrained system in which clients are connected to a wireless Access Point (AP). We choose video streaming as the application of interest using the case study of YouTube, since video has stringent network requirements and occupies a majority of Internet packets today \cite{ericsson-mobility-report}. The AP has a high and low priority queue.  Clients assigned to the high priority queue typically  experience a better QoS (higher bandwidth, lower latency etc.) when compared to the clients assigned to the low priority queue.  We assume that a fixed number of clients can be assigned to the high priority queue. The controller \textit{optimally} assigns clients to each of these queues at every decision period (DP; 10 seconds in our implementation). 

\subsection{System-wide Model}
\label{section:system,systemwide}
We consider a discrete time system where time is indexed by $t \in \{0,1,...\}$.  At each DP ($t=0,1,2..$) the controller observes the \textit{state} of the system and assigns clients to queues based on a \textit{policy}. This problem can be modeled as a  Markov Decision Process (MDP) consisting of an \textit{Environment} that produces \textit{states} and \textit{rewards} and an \textit{Agent} that takes \textit{actions}.

\textbf{Environment:} The environment is composed of the AP and clients.  Let $\mathcal{C}$ denote the set of all clients.

\textbf{Client State:}  Each client keeps track of its state which consists of its current buffer (the number of seconds of video that it has buffered up), the number of stalls it has experienced (i.e., the number of times that it has experienced a break in playout and consequent re-buffering), and its current QoE ( a number in $[1,5]$ that represents user satisfaction, with 5 being the best). Let $s^c_t$ denote the state of client $c$ at time $t$.  
 
\textbf{System State:} The state of the system is the union of the states of all clients. Let $s_t$ denote the state of the system and $\mathbb{S}$ denote the set of all possible system states,
\begin{align*}
s_t^c &=\textit{[Current Buffer State, Stall  Information,} \\
&\textit{ Current QoE]  } \forall c \in \mathcal{C} \\
s_t&=\left[\cup_{\forall c \in \mathcal{C}} s_t^c\right]
\end{align*}

\textbf{Agent:} The controller is the agent, that assigns flows to queue every decision period based on a policy.  The queue assignment performed by  the controller at time $t$ is called action $a_t \in \mathcal{A}$. Let $a_t^c$ denote the assignment of client $c$ at time $t$,
$$a_t=[\cup_{\forall c \in \mathcal{C}} a_t^c] $$

\textbf{Reward:} The reward $R(s_t,a_t)$ obtained by taking action $a_t$ at state $s_t$ is the expected QoE of all clients in state $s_{t+1}$.

\textbf{Policy:} The system-wide goal is to maximize the overall QoE.  This goal can be formulated as determining the optimal policy $\pi^*$, that maximizes the Bellman Optimality Equation 
\begin{equation}
\begin{aligned}
&\pi^*(s_t)=\\ 
&\text{argmax}_{a_t \in \mathcal{A}} (R(s_t,a_t)+ \gamma \sum_{s_{t+1} \in \mathbb{S}} \mathbb{P}(s_{t+1}|s_t,a_t)V^{*}(s_{t+1})),
\end{aligned}
\label{eq:bel}
\end{equation}
where $\gamma$ is the discount factor, $\mathbb{P}(s_{t+1}|s_t,a_t)$ is the \textit{system transition kernel} and $V^{*}(s_{t+1})$ is the optimal value of state $s_{t+1}$.

\subsection{Auction based Model}
\label{section:system,auctionbased}
We consider a market wherein clients bid for high priority service periodically. In each discrete time instant, a fixed number of clients $N$ are assigned to the high priority queue. Clients participate in an $(N+1)^{th}$ auction to compete for admission to the high priority queue. The $N$ winners who obtain high priority services will pay a price that is equal to the $(N+1)^{th}$ highest bid and the rest of the clients will be assigned to the low priority queue. We model the system in a Mean Field approach as described below,

{\bf{Bid}:} The bid submitted by the client in each auction is denoted by $b\in\mathcal{B}$, where $\mathcal{B}$ is a set containing discrete bid values.  The bids can be seen as the price each client $c$ is willing to pay to obtain high priority service.  Note that the human end user plays no role in selecting these bids.

{\bf{Bid Distribution}:} The clients must place their bid based on the beliefs of their competitors. We denote the assumed bid distribution in the market as $\rho$.

{\bf{Payment}:} The amount of transaction after each auction is denoted by $pay$. Note that $pay$ is a random variable that corresponds to the auction mechanism. In particular, the payment distribution in our system is exactly the distribution of the $(N+1)^{th}$ highest bid.

{\bf{Client Reward}:} Reward $R(s_{t}^{c},a_{t}^{c})$ resulting from  action $a_t^c$ at state $s_t^{c}$ is the expected QoE of client $c$ in state $s_{t+1}^c$.

\textbf{Client Transition Kernel} Let $\mathbb{P}(s_{t+1}^{c}|s_t^{c},a_t^{c})$ denote the client transition kernel. The action $a_t^{c}=win$ and $a_t^{c}=lose$ corresponds to assignment of client $c$ to the high and low priority queue respectively. Thus the probability of transitioning to state $s^{c}_{t+1}$ is jointly defined by the probability of winning the auction when bidding $b$, $p_{win}(b)$ and  $\mathbb{P}(s^c_{t+1}|s^c_t, a^{c}_t)$.

{\bf{Policy}:} From a single client's perspective, all his opponents are placing bids according to the same bid distribution, which is a public belief. Thus we formulate the policy of the corresponding MDP as follows, 
\begin{align}
&b^{*}(s^c_t)= 
\text{argmax}_{b\in\mathcal{B}} \bigg\{ p_{win}(b)\Big[R(s^c_{t+1}, a^{c}_t=win) - pay + \nonumber\\ 
&\sum_{s^c_{t+1}}\mathbb{P}(s^c_{t+1}|s^c_t,a^{c}_t= win)\gamma v(s^c_{t+1})\Big] +\nonumber\\
&  (1- p_{win}(b))\Big[R(s^c_{t+1},a_t^{c}=lose) + \nonumber\\
&\sum_{s^c_{t+1}}\mathbb{P}(s^c_{t+1}|s^c_t,a_t^{c}= lose)\gamma v(s^c_{t+1})\Big] \bigg\}
\label{eq:val}
\end{align}

\subsection{Measuring QoE for Video Streaming}
\label{section:system,qoe}

Various models validated by extensive human experiments 
identify the relation between video events and subjective user perception (QoE).  These models are based on video stalling events if there is no rate adaptation.  Since our goal is to support high video resolution, we fix the video resolution to prevent rate adaption. We choose the Delivery Quality Score (DQS) \cite{7025402} as our QoE model. 
\begin{figure}[!htbp]
\vspace{-0.15in}
\begin{center}
\includegraphics[width=2.3in]{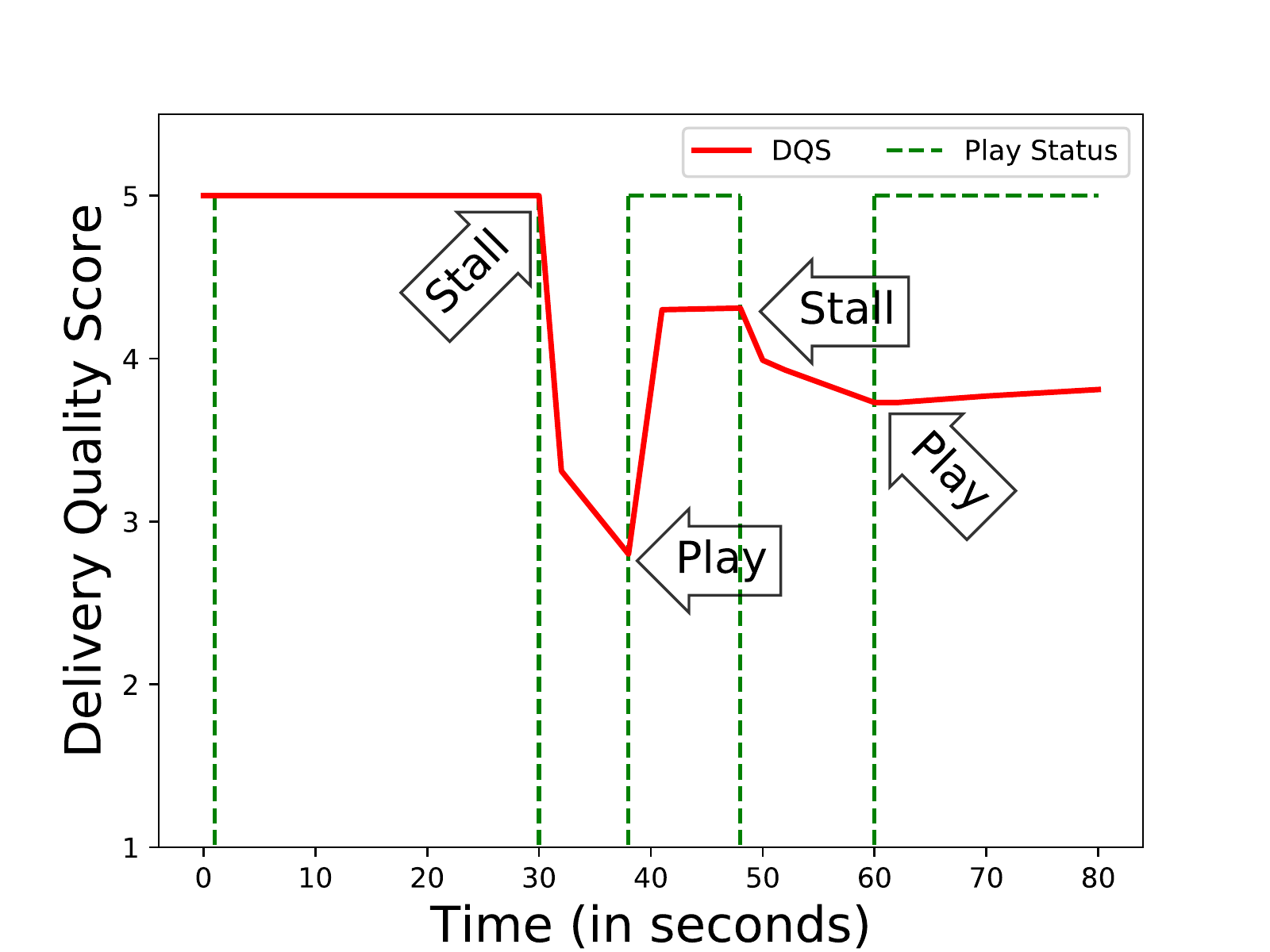}
\caption{Evolution of DQS}
\label{fig:dqs_evol}
\end{center}
\vspace{-0.215in}
\end{figure}

The DQS model takes various factor into account (such as stall duration and number of stalls) to measure the impact of a stall on QoE. Figure \ref{fig:dqs_evol} \cite{bhattacharyya2019qflow} shows the evolution of QoE over the duration of a video. Observe that the impact of the first stall event on QoE is large when compared to subsequent stall events. The change in perceived QoE is captured by a function which is is a combination of raised cosine and ramp functions. Recovery of QoE from each stall event becomes progressively harder. Note that DQS has been validated using 183 videos and 53 human subjects \cite{7025402}, and we do not repeat the user validation experiments.

\section{System Architecture}
\label{architecture}

\begin{figure}[htbp]
\vspace{-0.15in}
\begin{center}
\includegraphics[width=3.6in]{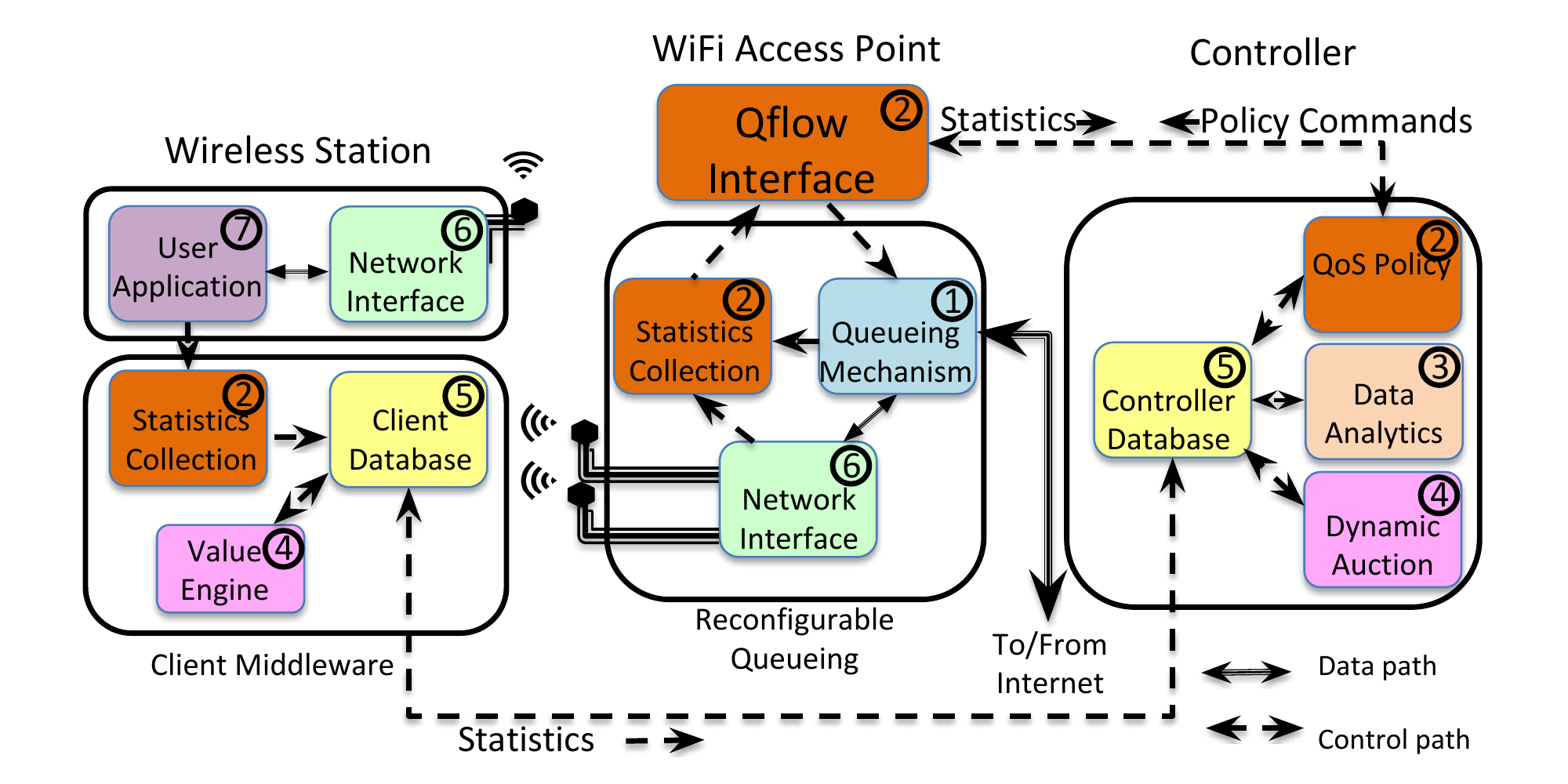}
\caption{FlowBazaar Architecture.}
\label{fig:architecture}
\end{center}
\vspace{-0.15in}
\end{figure}

The architecture of FlowBazaar, illustrated in Figure~\ref{fig:architecture}, is an extension to QFlow \cite{bhattacharyya2019qflow}.  
The three main units in our system are an off-the-shelf WiFi Access Point running OpenWRT, multiple wireless stations installed with custom middleware, and a centralized controller. The units contain different functional components which are shown in a color coded manner. These components have functionalities pertaining to packet mechanisms, QoS policy, application QoE, and end-user value, which we overview below. Tying together the units are a Controller Database in which we log all events, and a smaller Client Database at each station that obtains a subset of the data that it needs for decision making, both shown as yellow tiles. The components related to Network Interface and User Application are unaware of the rest of the system.

\textbf{Queueing Mechanisms (blue tiles):}
We create multiple MAC Layer queues, and apply different packet scheduling mechanisms over them. The impact of these mechanisms on flows is seen on the resultant QoS statistics at the queue level, which in turn affects application performance. Flows in high priority queues experience much better performance when compared to those in low priority queues.

\textbf{QoS Policy (orange tiles):}
The centralized controller makes policy decisions (assignment of flows to queues) which are communicated to the Access Point using the OpenFlow protocol. We reuse a custom message format described in \cite{bhattacharyya2019qflow} to send MAC Layer commands. The Access Point is installed with SoftStack, which interprets the received messages and implements the policies selected by the controller. The Statistics Collection component at the Access Point periodically collects network connection statistics like throughput, drop rates and RTT and sends them back to the controller in a predefined message format using the OpenFlow protocol.

\textbf{Application QoE (beige tiles):}
A smart middleware layer at clients is used to interface with our system. It retrieves the state of the foreground application and translates it into the impact on perceived QoE using the DQS model.  This layer also ensures that the system components remain invisible to the application and end user.

\textbf{End-user Value (pink tiles):}
In every decision period (10 seconds), clients are offered high and low priority service under an $(N+1)^{th}$ price auction, with the top $N$ bidders being admitted to the high priority queue.  
The Value Engine retrieves the application state from the Client Database and the statistics of the current market conditions (bid distribution) from the Controller Database. On receiving both, it is responsible for running the Value Iteration to obtain the optimal Value Function, which determines what the value of winning and losing would be. The Value Engine then sends an appropriate bid over to the Controller Database. The Dynamic Auction module conducts the auction, and the the resulting assignment remains in place for $10$ seconds. 

\textbf{Order of Interactions:}
The Client Middleware uses the client application state to determine the value function using the using the Value Engine and places a bid accordingly.  Bids are sent to the  the Controller, which conducts the auction, and sends the policy decision using OpenFlow messages that are sent to the Access Point.  SoftStack interprets and implements these policy decisions.  The cycle repeats every 10 seconds.  Client and queue statistics are collected every second.

\section{Policy Design}
\label{section:policy}

\subsection{Transition Kernel of an Individual Client}
\label{section:policy,clientkernel}
To determine optimal policies, we first need to compute the transition kernel of an individual client, ie., we need to determine $\mathbb{P}(s^c_{t+1}|s^{c}_t,a_t^c)$. We do this as follows,   
\begin{enumerate}
\item We generate state ($s_t^c$), action ($a_t^c$) and next state ($s_{t+1}^c$) tuples for clients by running the system described in Section \ref{architecture} for a duration of 10 hours under vanilla, round robin and greedy buffer (discussed in Section \ref{section: experiments}) policy.
\item The state $s_t^c$ is a 3 dimensional vector (Buffer, Stall, QoE) consisting of continuous values, thus we discretize the state space. 
We encode the discretized state space to obtain a label for each state. Let $NSB$ and $NQB$ denote the number of stall and QoE bins respectively,
$$s_t^c = \text{Buffer}\times \text{NSB} \times \text{NQB } + \text{QoE}\times \text{NSB } + \text{Stall} $$  
\item We fit an empirical distribution over the discretized data to obtain the transition kernel.        
\end{enumerate}

\subsection{Optimal Policy for System-wide Model}
\label{section:policy,systemwidemodel}
To obtain the optimal policy for the system-wide model, we have to  first determine the system transition kernel i.e., given the current state $s_t$ of the system and the action taken $a_t$, we find the transition probabilities to the next states $s_{t+1}$.  Given the transition kernel of the system, we can use  policy or value iteration to solve for the optimal policy $\pi^{*}$.  The system-wide approach is particularly interesting because of its special structure, since the state transitions of a client given its current state and action are independent of the states and actions of other clients in the system.  In other words,
$$ \mathbb{P}(s_{t+1}|a_t,s_t)=\prod_{\forall c \in \mathcal{C}}\mathbb{P}(s^c_{t+1}|a_t^c,s^{c}_t)$$
It must also be noted that the state transitions of all clients in the system given their current states and actions are identical.  Thus, we can determine the transition kernel of the system using the transition kernel of each individual client. \\
Directly determining the system transition kernel and solving it to obtain the optimal policy is intractable due to the large state space of the system. Consider a system with $N$ clients (the state space is an $N$ dimensional vector, with each dimension corresponding to the state of a client ), where each client has a state space of the order of $10^3$. The state space of the system ($\mathcal{S}$) is of the order of $10^{3N}$ which is very large. To combat the problem of state space explosion, we take advantage of the structure of the problem and identify the \textit{popular states} of the system ($\mathcal{S}_p$), and approximate all the other states to the closest popular state under the $L^2$ norm. \\
To obtain the system transition kernel, we empirically fit a distribution over the transitions generated for each state in $\mathcal{S}_p$ under each action in $\mathcal{A}$ using the transition kernel of an individual client (Section \ref{section:policy,clientkernel}). If the transitions generated are outside $\mathcal{S}_p$ we approximate it with the state closest in $\mathcal{S}_p$  under the $L^2$ norm. Once the system transition kernel is obtained, we run value iteration solve (\ref{eq:bel}) to obtain the optimal policy. It must be noted that the reward obtained by taking action $a_t$ in state $s_t$ is the expected QoE of state $s_{t+1}$.


\subsection{Optimal Policy for Auction-based Model}
\label{section:policy,auctionbased}
Using the transition kernel of a client (Section \ref{section:policy,clientkernel}), we use Value Iteration to solve  (\ref{eq:val}) and obtain the optimal value function for an individual client. This approach also provides a map between state of a client and its bid, which is subsequently used in the $(N+1)^{th}$ price  auction. Since it is known that $(N+1)^{th}$-price auction promotes truth telling, a client participating in such an auction makes a bid which reflects its true value. The Auction Agent receives the bids from all clients, conducts the $(N+1)^{th}$-price auction and performs the assignment on the basis of the result.

\subsection{Index Policy}
\label{section:policy,index}
The solution to (\ref{eq:val}) results in a \textit{value} for each state  $s_t^{c}$ of the client.  We can order states in increasing order of value, and associate each state with an \emph{index,} which is its position in the order.    Then these indices can be used to directly decide which clients to prioritise, and we call this as an \emph{index policy.}  Now, given the indices corresponding to a system with $N$ clients, it would save computational effort if we could use the same indices for a system with $M < N$ clients, by simply setting indices of non-existent clients to  0.  The question is whether the indices for a system with 6 clients are consistent with (for example) one that has 3 clients?
\begin{figure}[htbp]
\vspace{-0.1in}
\begin{center}
\includegraphics[width=3in]{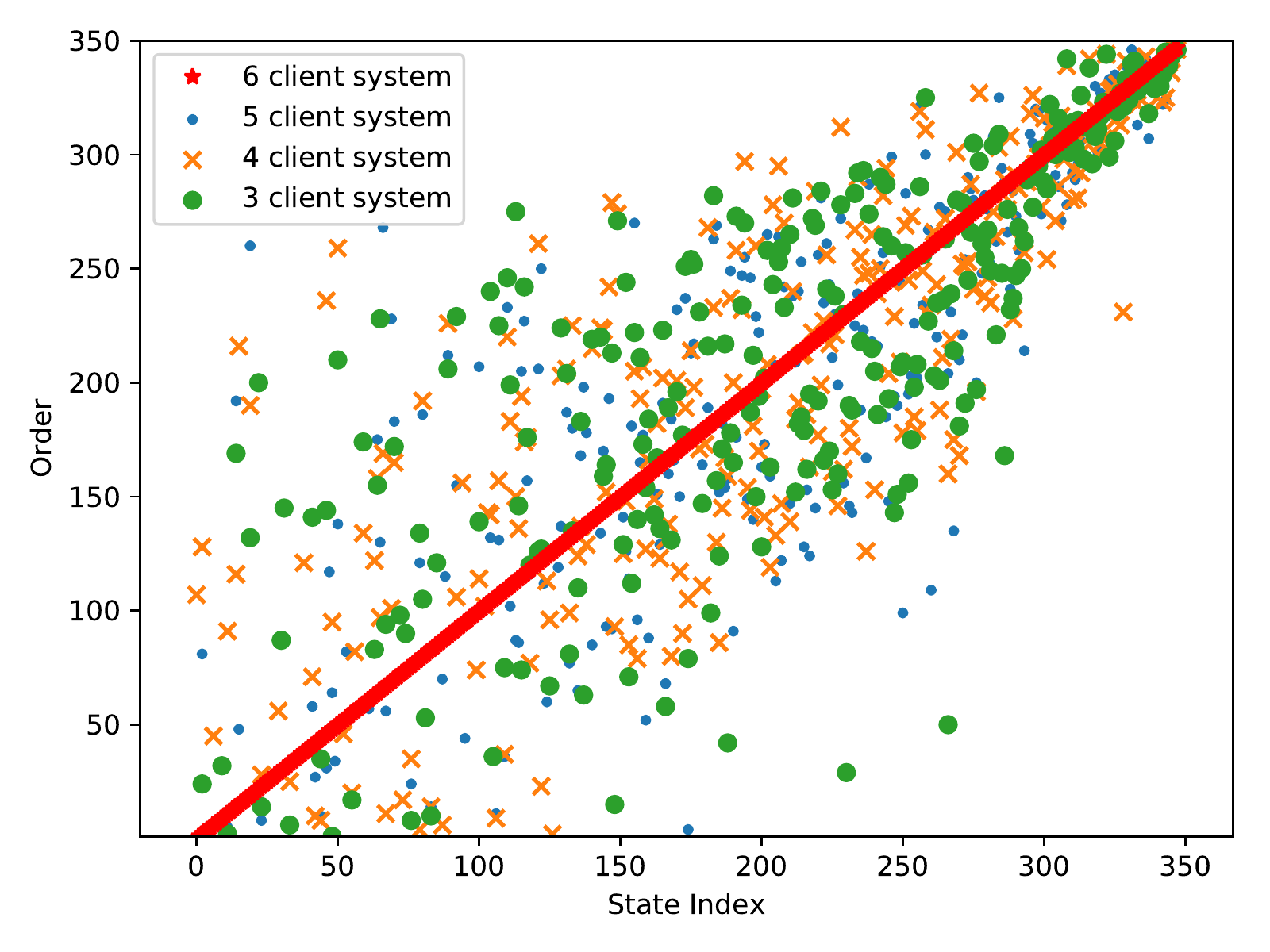}
\caption{Ordering of states in different client configurations.}
\label{fig:index}
\end{center}
\vspace{-0.2in}
\end{figure}


We experimentally determined the values for different numbers of clients, and calculated the state indices in each case.   The comparison of the orderings for different client configurations (6, 5, 4, 3 clients) is shown in Figure \ref{fig:index}, using the ordering for 6 clients as the base ordering (we have not shown several hundred states that have an index of 0).  We observe that the relative ordering of most of the high value states is consistent across configurations, 
which indicates that an index policy for 6 clients is likely to perform well for one with fewer clients.  



\subsection{Dynamic number of Clients and varying Channel}
\label{section:policy,dynamic}
In the previous subsections we assumed that the number of clients and channel state of the system were static. To deal with a dynamic environment, ie., varying number of clients and channel states (described in Section \ref{section: experiments}), we first obtain the optimal policy for all the different scenarios using the static approach described in section. We then construct a composite controller which chooses the appropriate policy based on the environment. This approach works well in practice since the time scale in which the environment changes is larger than the decision period.      

\section{Evaluation}
\label{section: experiments}

\begin{figure*}[htbp]
\centering
\begin{minipage}{.32\textwidth}
\centering
\includegraphics[width=1\columnwidth]{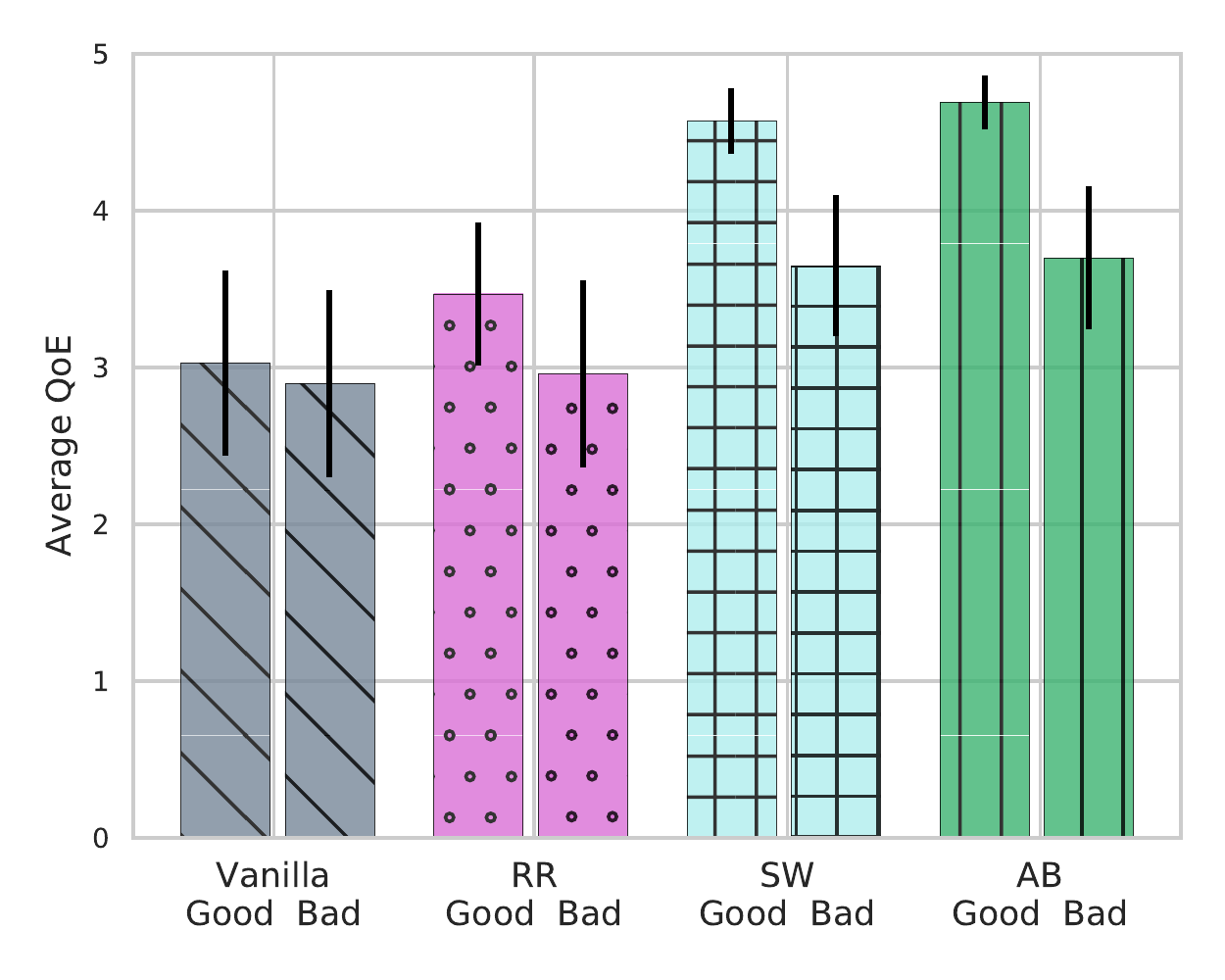}
\caption{Comparison of average QoE }
\label{fig:qoe_comp}
\end{minipage}\hfill
\begin{minipage}{.32\textwidth}
\centering
\includegraphics[width=1\columnwidth]{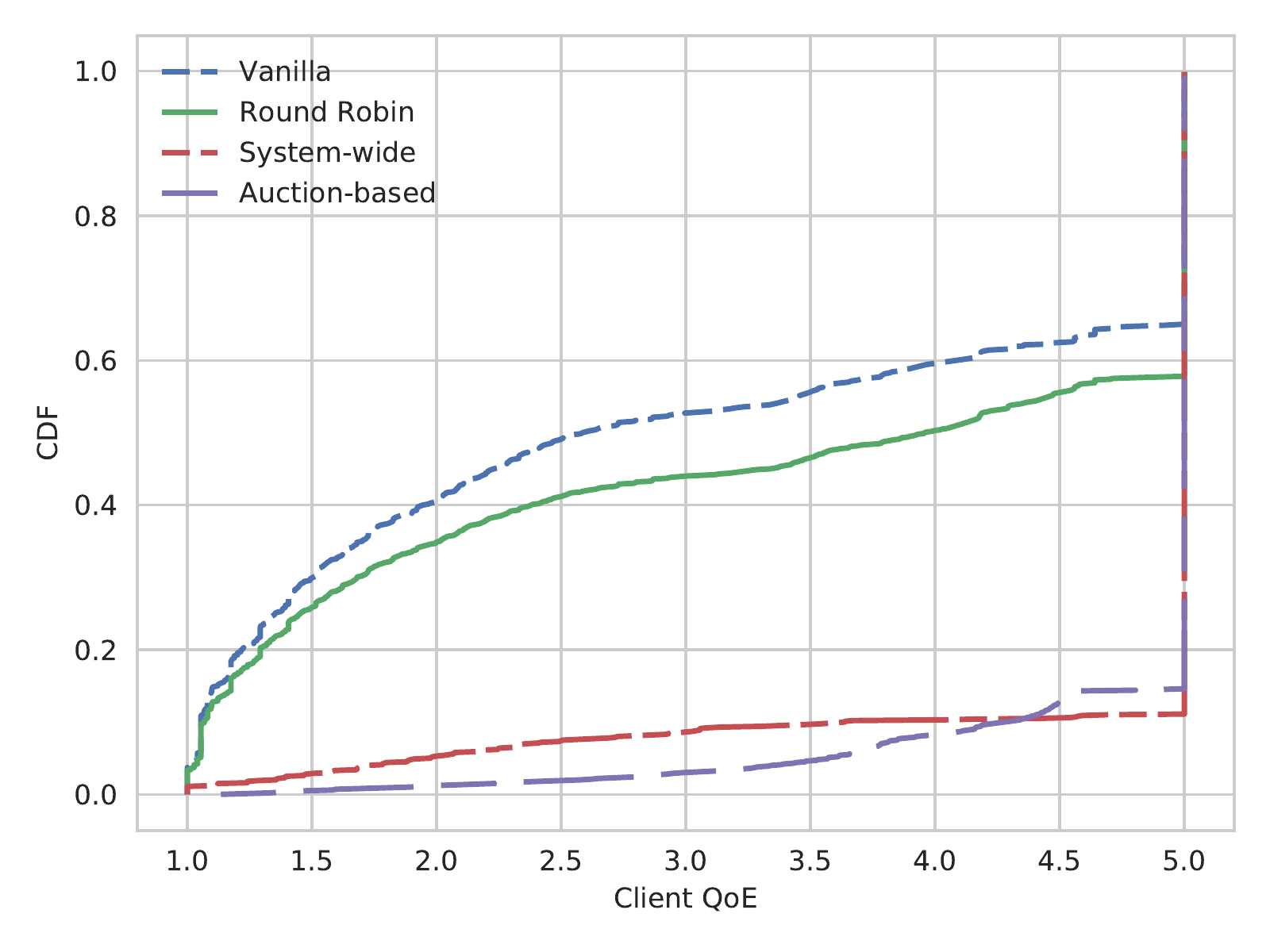}
\caption{Client QoE comparison for Good Channel}
\label{fig:qoe_cdf}
\end{minipage}\hfill
\begin{minipage}{.32\textwidth}
\centering
\includegraphics[width=1\columnwidth]{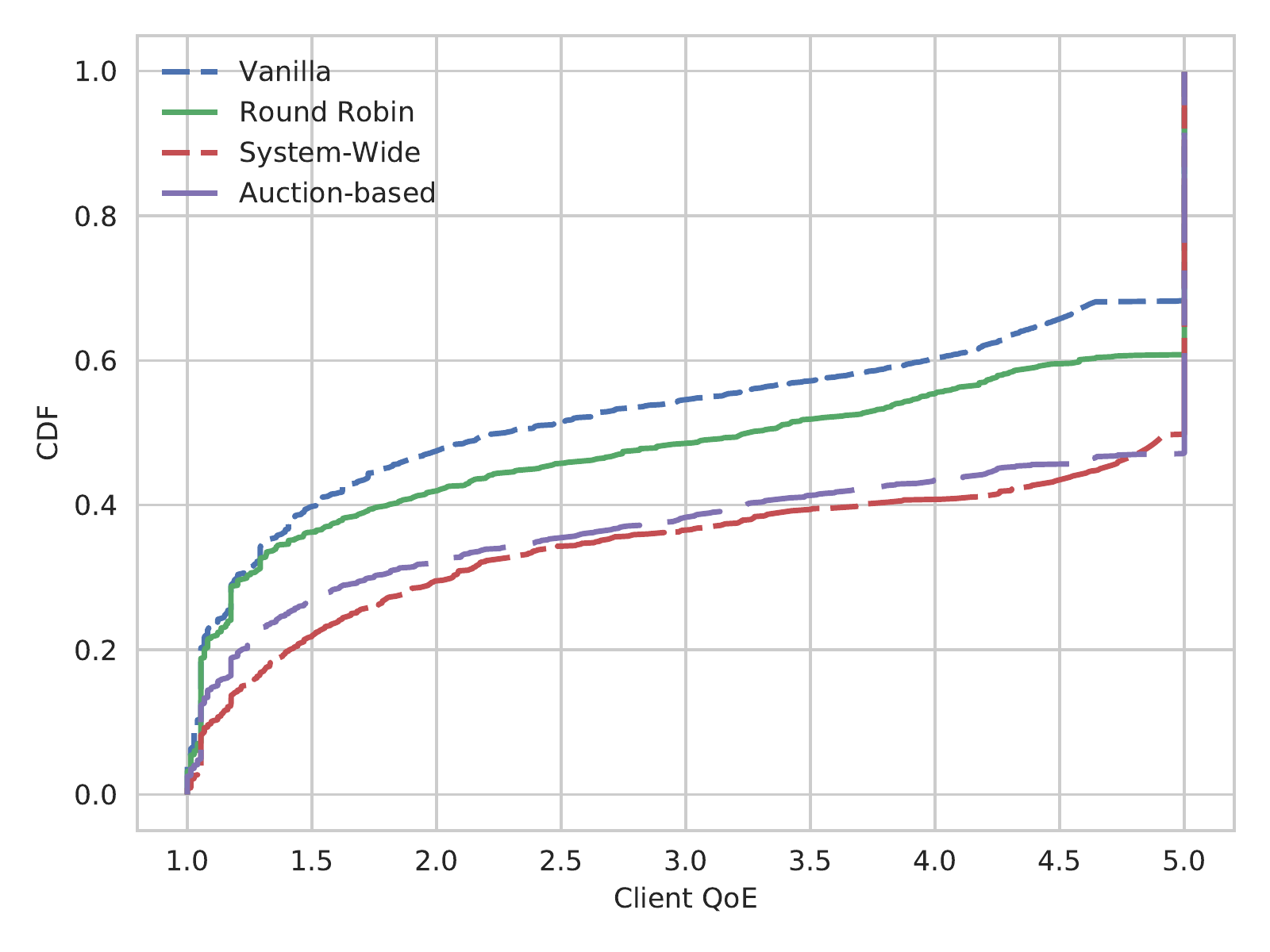}
\caption{Client QoE comparison for Bad Channel}
\label{fig:qoe_cdf_bad}
\end{minipage}
\begin{minipage}{.32\textwidth}
\centering
\includegraphics[width=1\columnwidth]{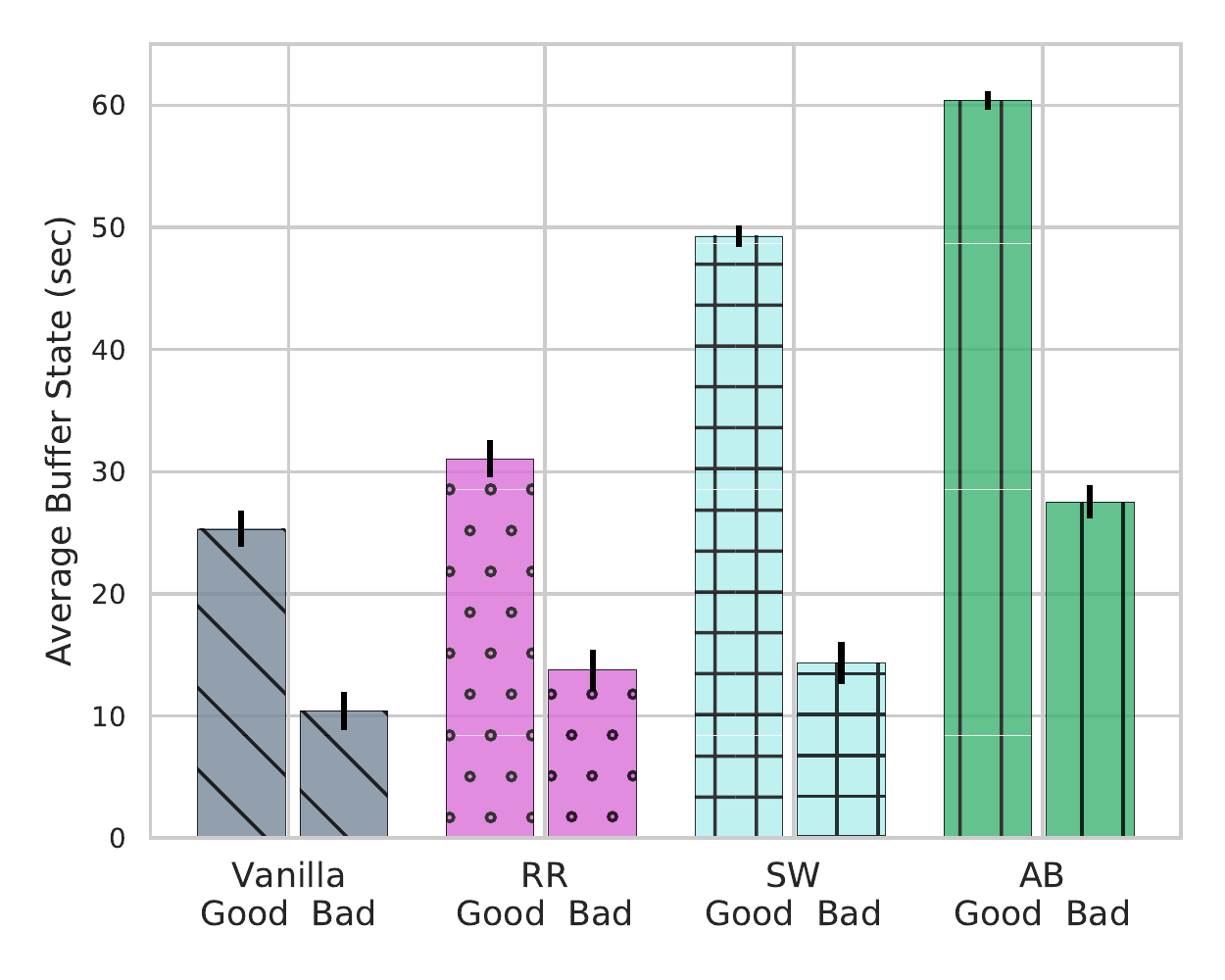}
\caption{Comparison of average Buffer }
\label{fig:buf_comp}
\end{minipage}\hfill
\begin{minipage}{.32\textwidth}
\centering
\includegraphics[width=1\columnwidth]{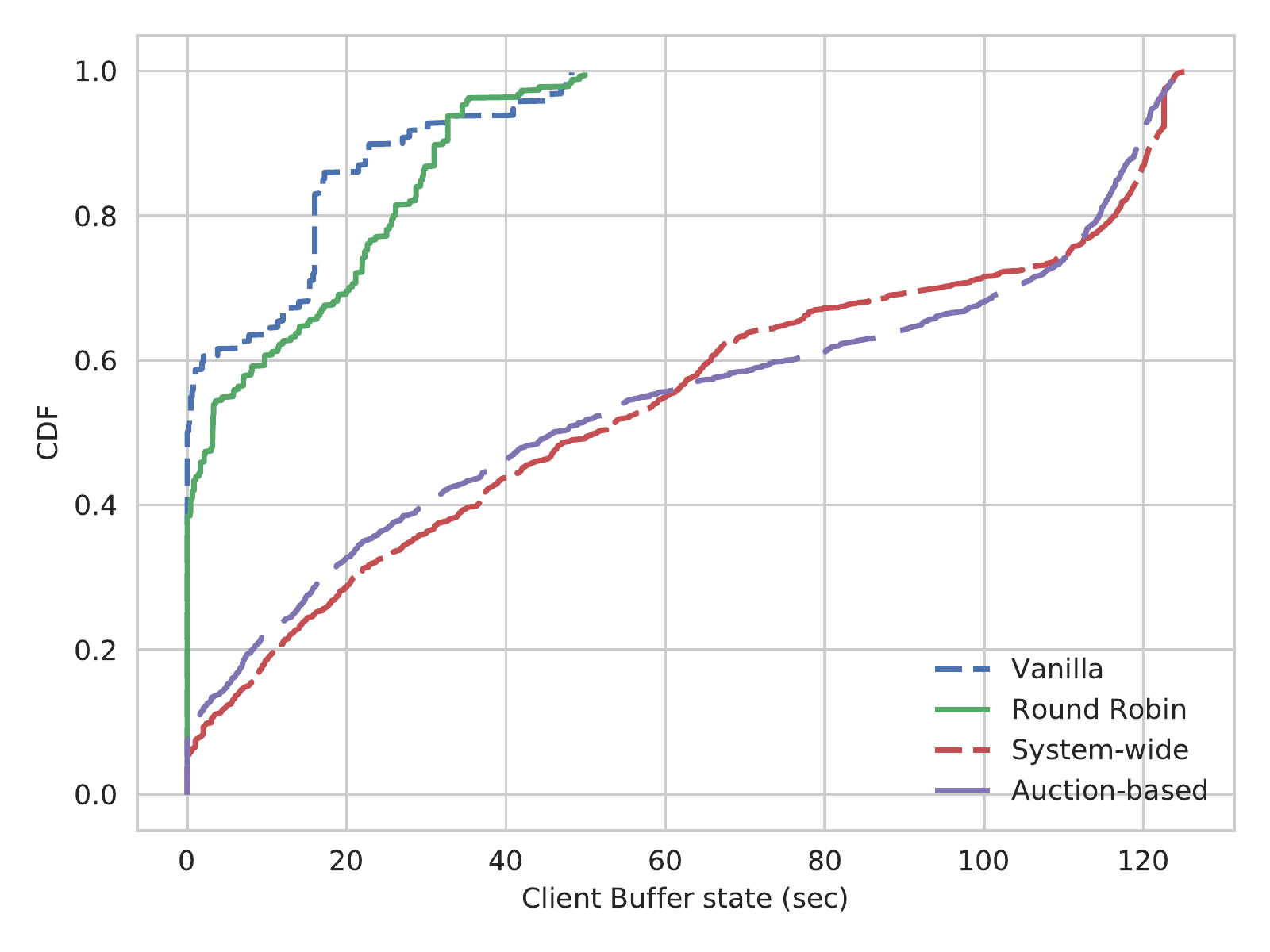}
\caption{Client Buffer comparison for Good Channel}
\label{fig:buf_cdf}
\end{minipage}\hfill
\begin{minipage}{.32\textwidth}
\centering
\includegraphics[width=1\columnwidth]{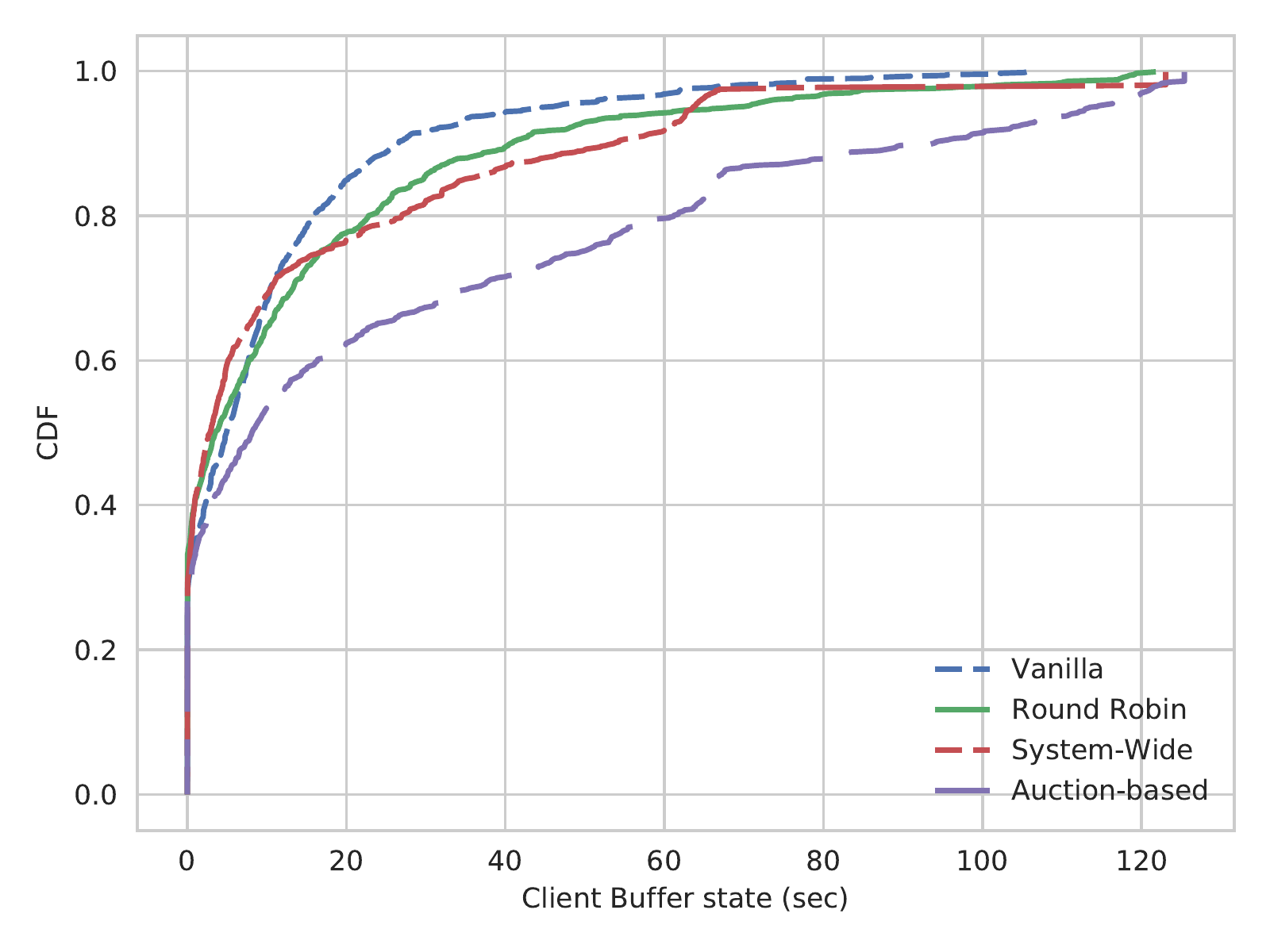}
\caption{Client Buffer comparison for Bad Channel}
\label{fig:buf_cdf_bad}
\end{minipage}
\vspace{-0.1in}
\end{figure*}

We ran a series of experiments to evaluate the performance of the described policies on a testbed hosting multiple YouTube sessions. A WiFi router installed with SoftStack is used as the Access Point and three Intel NUCs are used to simulate up to 9 clients (YouTube sessions) for our experiments. Although each YouTube session can be associated with multiple TCP flows, all such flows are treated identically.  Each of the NUCs is configured with an i7 processor and 8 GBs of memory, and is  powerful enough to host multiple traffic intensive sessions simultaneously. Ubuntu Operating system is installed on the NUCs, which makes it easier to measure session specific information such as ports used by an application, play/load progress, bitrate and stall information. This information is collected every second and written to the centralized database for ease of sharing.

Since we are interested in studying the behavior of queues in routers, we create two bins of downlink queues, each containing a high priority and a low priority queue. The motivation behind the creation of two separate bins is to ensure that clients having similar signal strengths are eligible for the same bin and hence, do not adversely affect the performance of client with better signal strengths. Hence, we have a $Good$ bin for clients with high signal strengths and a $Bad$ bin for those who have low signal strengths. These bins are attached to the WiFi interface of our AP using \emph{tc}. We allocate a higher bandwidth to each of the high priority queues using hierarchical token bucket queueing discipline. This is done so that clients assigned to these (high priority) queues experience better service than those in the other (low priority) queues. For the current scope of experiments, we set the admission limit of the high priority queues to two clients. We also create a default queue to accommodate any background traffic. 

\textbf{Emulation of bad network conditions}
Since we have a fixed number of NUCs to host a total of upto 9 sessions, we decided to emulate bad channel by reducing the throughput and increasing the latency and loss rates of the queues in the $Bad$ bin as compared to those in the $Good$ bin. We ran several hours of experiments with clients having low signal strengths to come up with these heuristics for emulating a bad channel. This enables us to mimic varying network conditions by dynamically assigning the sessions hosted on the NUCs to either the $Good$ or the $Bad$ bin.

Given this setup, the control problem is to determine the assignment of sessions to queues under varying channels.

\subsection{Policies}
We compare the \textbf{System-wide} (SW) policy described in Section \ref{section:system,systemwide} and the \textbf{Auction-based} (AB) policy described in Section \ref{section:system,auctionbased} with two other policies for determining the assignment of sessions to queues.

{\bf{Vanilla}:} This baseline scenario consists of a single queue which treats all clients equally. The queue is allocated the total bandwidth of the queues used for the other policies.

{\bf{Round Robin}:} The Round Robin (RR) policy assigns clients to the high priority queue in a cyclic manner.  It is simple, easy to implement, work-conserving and starvation-free. But it might promote the wrong clients (like clients who have stalled multiple times) to the high priority queue instead of those who might benefit much more from the assignment.

\subsection{Static Network Configuration}

We run the first set of experiments in a static configuration consisting of 6 clients hosted on three NUCs under both channel conditions.  For the $Good$ channel scenario, the total bandwidth allocation of the two downlink queues is set such that simultaneous playback of all the YouTube sessions cannot not be supported at HD (1080p), whereas for the $Bad$ channel scenario, we further reduce the bandwidth and add latency and loss. 
Figure \ref{fig:qoe_comp} shows the comparison of the average QoE achieved by the policies under the different channel conditions. It is evident from the figure that the System-wide and the Auction-based policies perform much better than the other policies.  
Interestingly, the Auction outperforms the System-wide policy.  We believe that this is due to the coarse quantization of the state space.  System-wide is worse affected by this due to the fact that 6 clients together are considered in the sate, whereas in Auction only 1 client is part of the (marginal) state.  Hence, we believe that value identification is more accurate in the Auction case.
 The difference in achieved performance for the different policies becomes more clear in the comparison of the CDFs of the client QoE under $Good$ and $Bad$ channel conditions in Figures \ref{fig:qoe_cdf} and \ref{fig:qoe_cdf_bad}. 
For example, we can observe from Figure \ref{fig:qoe_cdf} that the System-wide and the Auction-based policies are able to provide a QoE of 5 for almost 90 and 85\% of the time for all clients, whereas it is only about 40\% of the time for Round Robin. 
We can observe that this gap decreases in the $Bad$ channel scenario, but the System-wide and the Auction-based policies still achieve higher QoE for the clients.

\begin{figure*}[htbp]
\centering
\begin{minipage}{.32\textwidth}
\centering
\includegraphics[width=1\columnwidth]{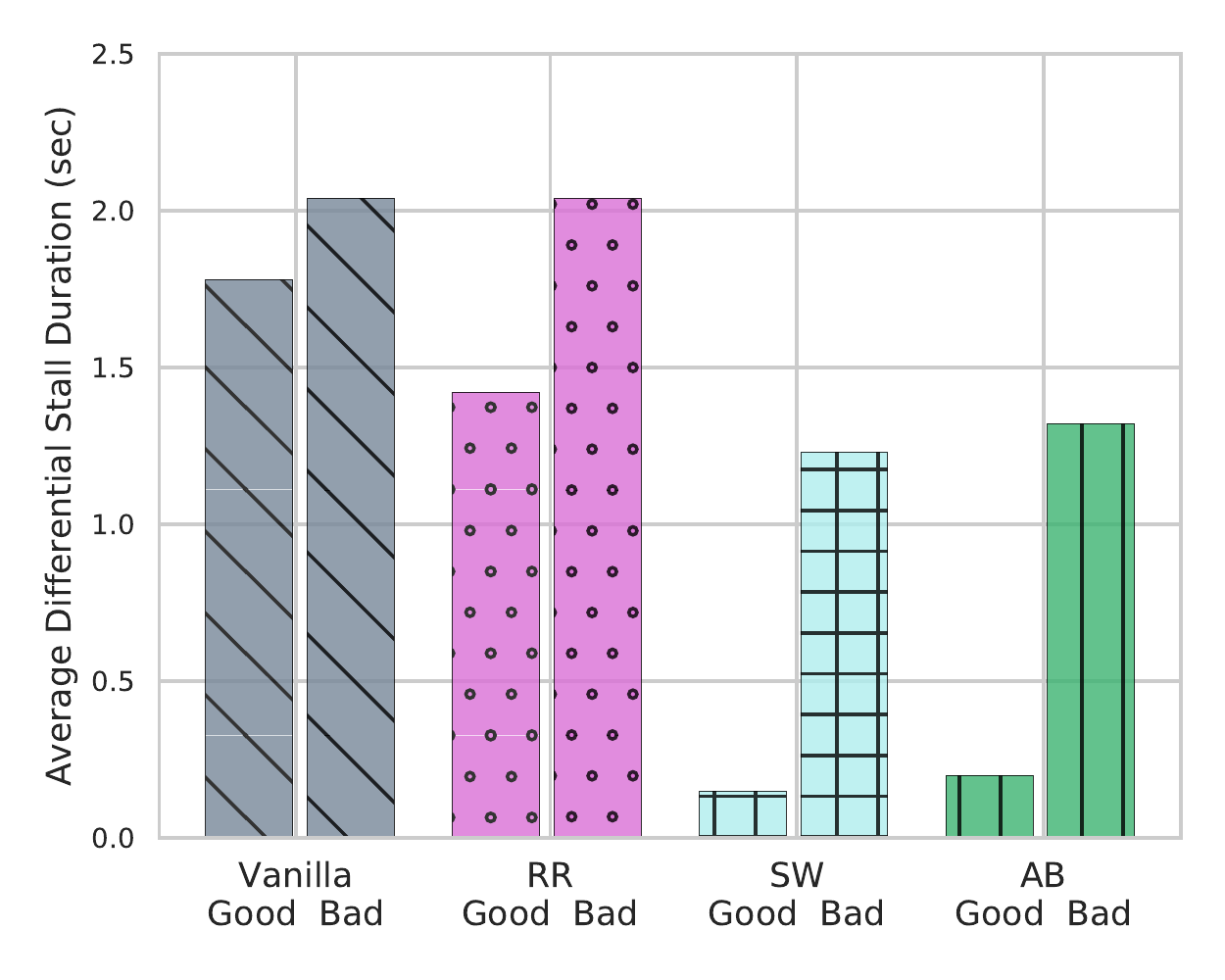}
\caption{Comparison of average stall duration }
\label{fig:stall_comp}
\end{minipage}\hfill
\begin{minipage}{.32\textwidth}
\centering
\includegraphics[width=1\columnwidth]{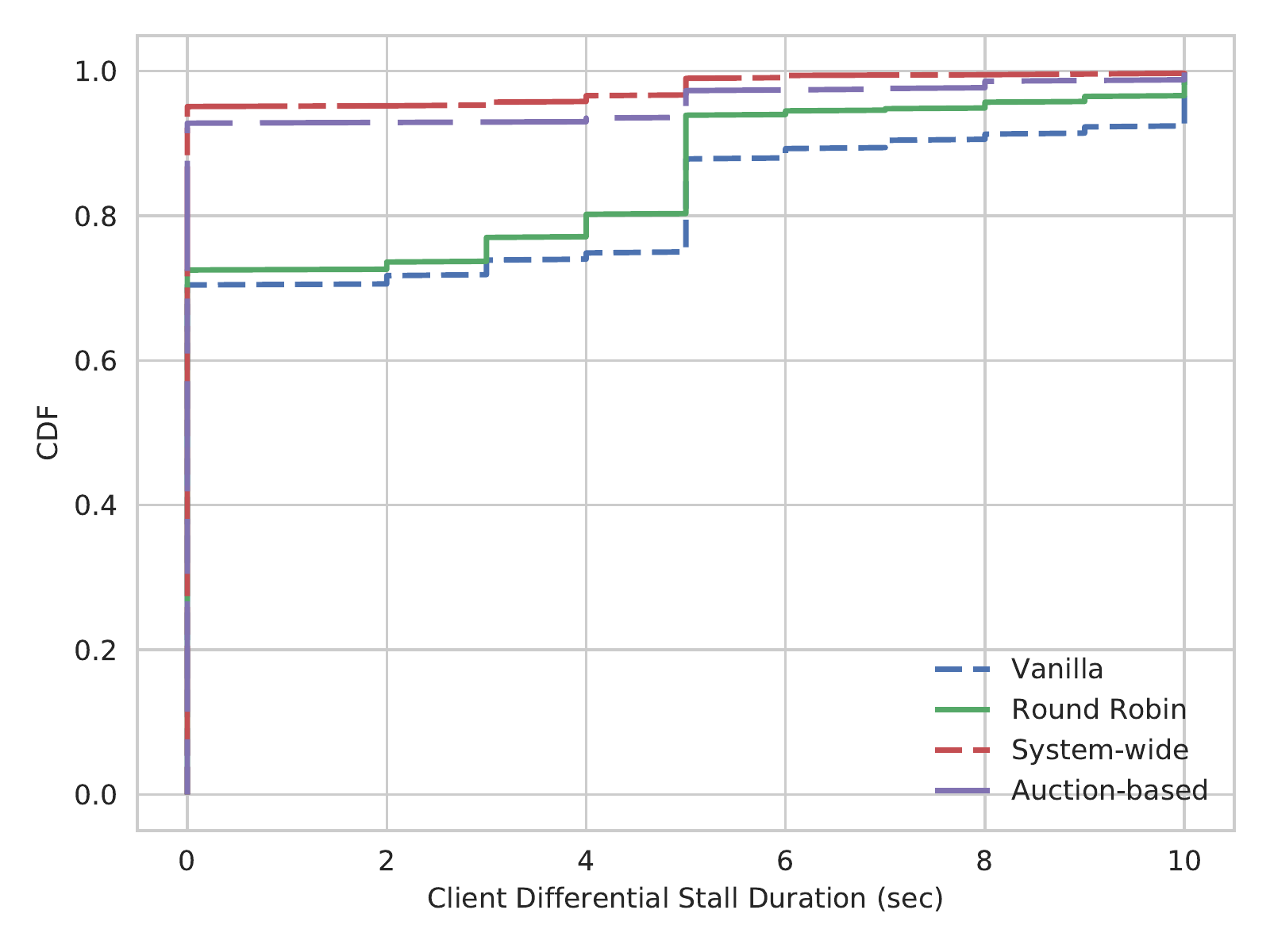}
\caption{Stall duration comparison for Good Channel}
\label{fig:stall_cdf}
\end{minipage}\hfill
\begin{minipage}{.32\textwidth}
\centering
\includegraphics[width=1\columnwidth]{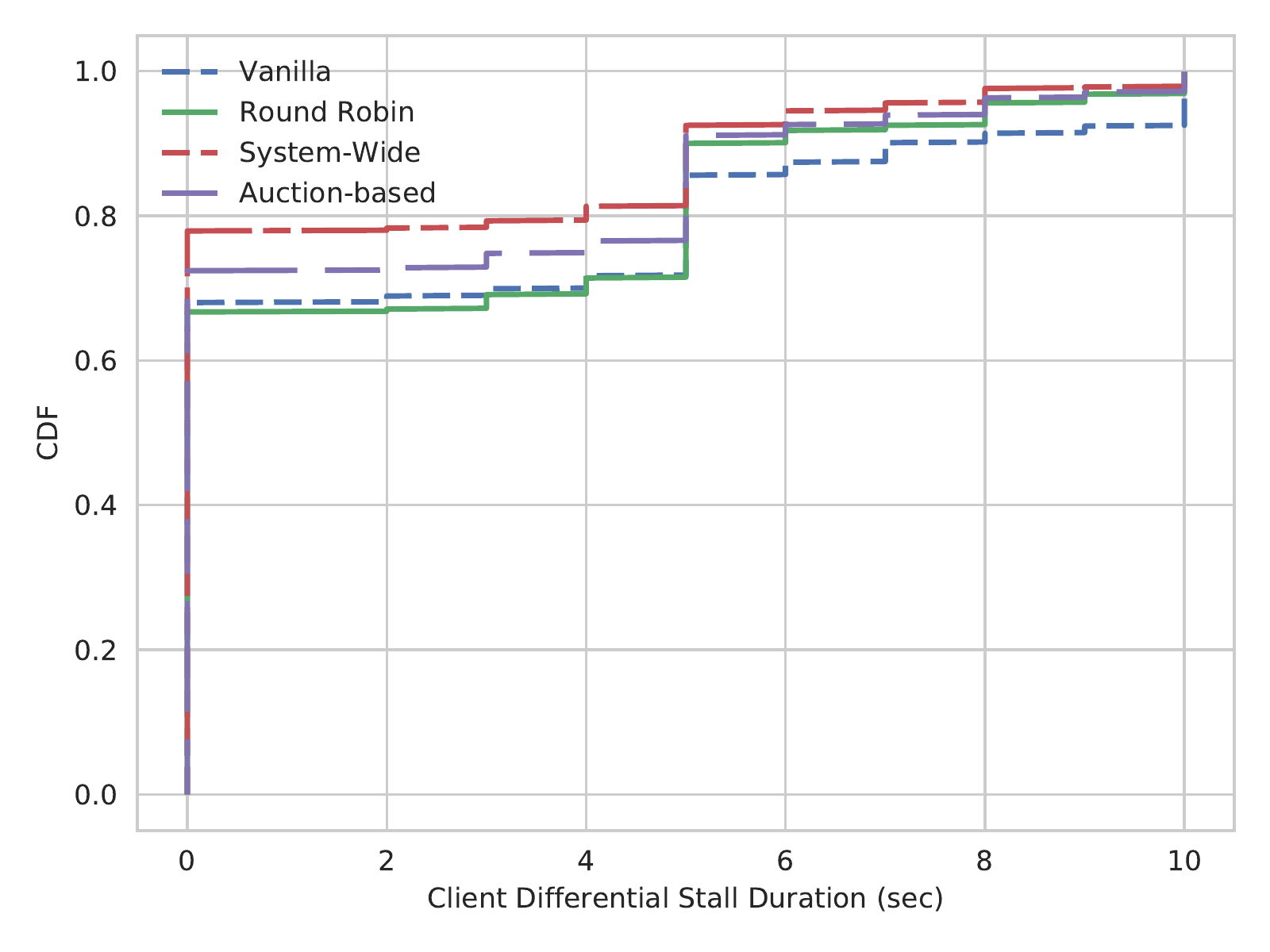}
\caption{Stall duration comparison for Bad Channel}
\label{fig:stall_cdf_bad}
\end{minipage}
\begin{minipage}{.32\textwidth}
\centering
\includegraphics[width=1\columnwidth]{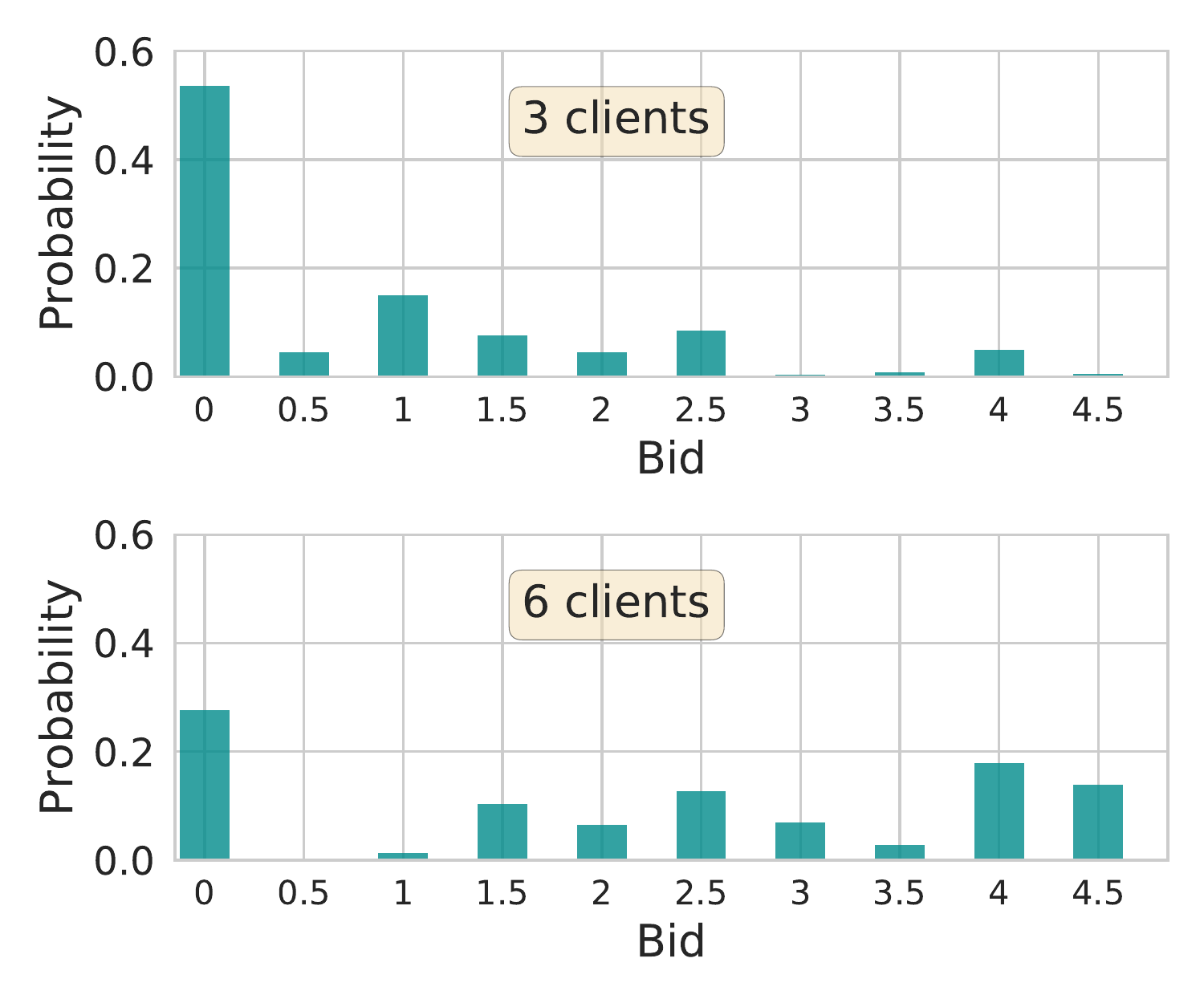}
\caption{Bid distribution for 6 and 3 client configurations}
\label{fig:bid_dist}
\end{minipage}
\begin{minipage}{.32\textwidth}
\centering
\includegraphics[width=1\columnwidth]{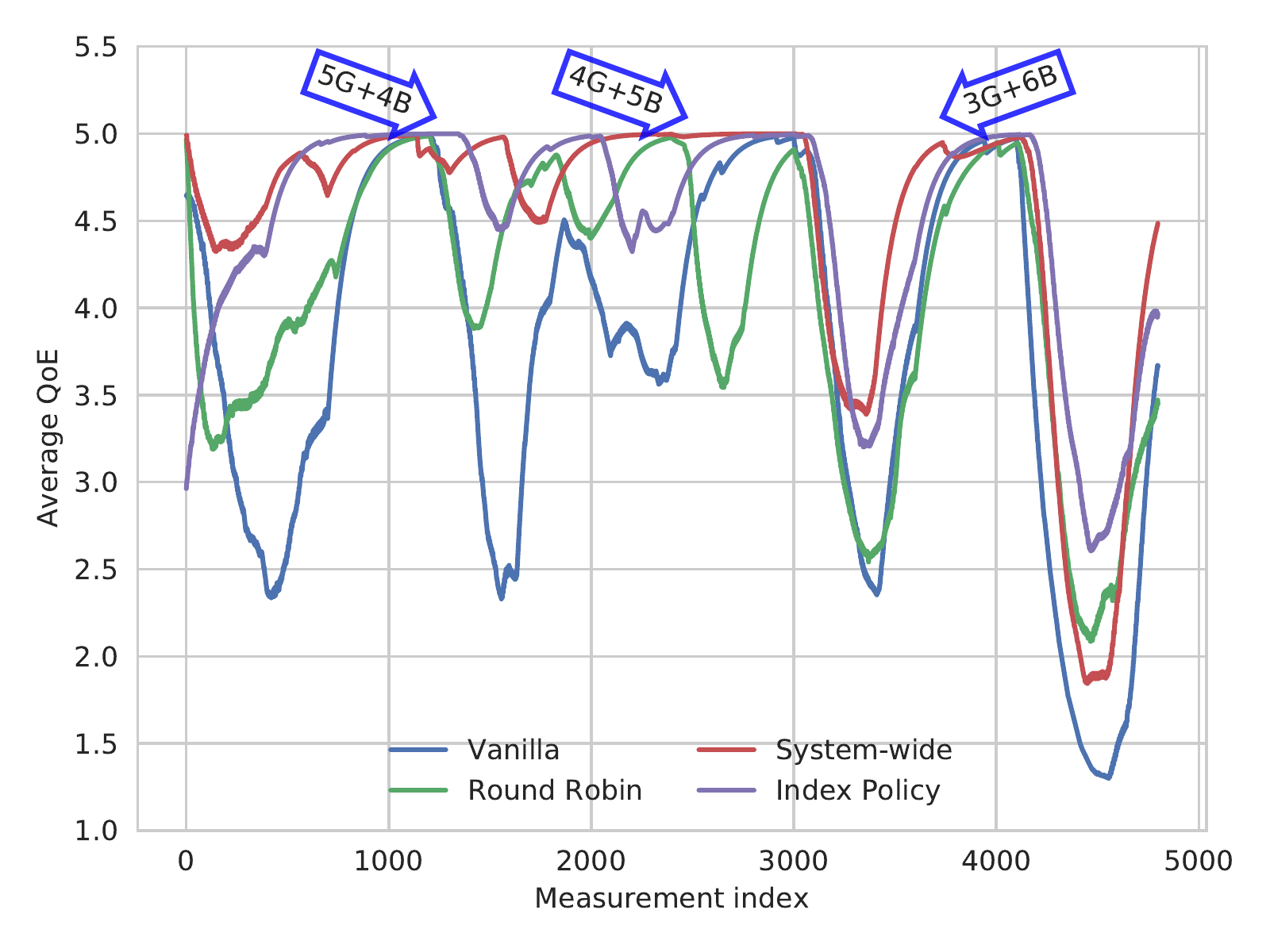}
\caption{Evolution of QoE dynamic clients in variable channel}
\label{fig:qoe_evol}
\end{minipage}\hfill
\begin{minipage}{.32\textwidth}
\centering
\includegraphics[width=1\columnwidth]{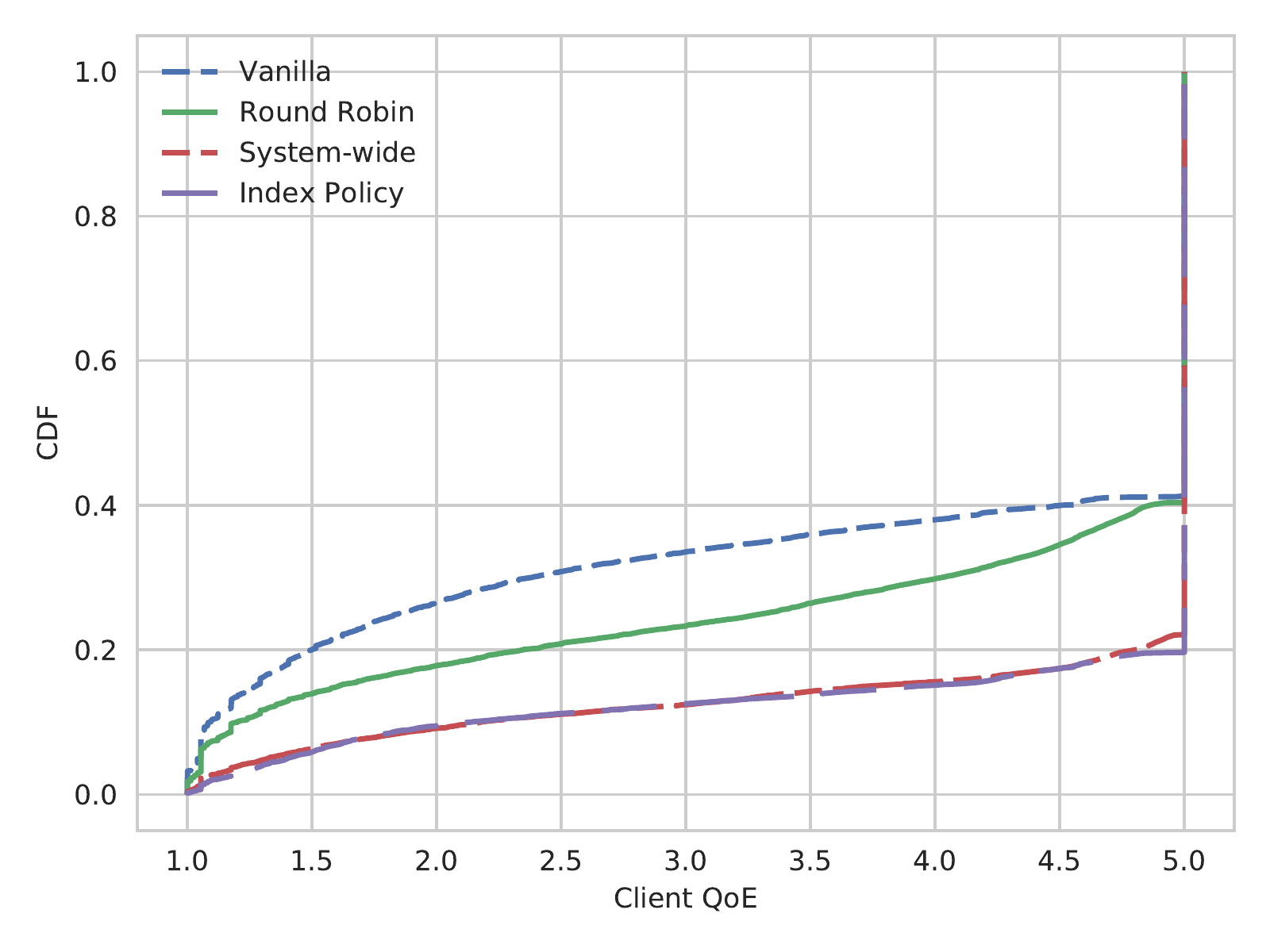}
\caption{Comparison of client QoE CDF for dynamic clients in variable channel}
\label{fig:qoe_channel_cdf}
\end{minipage}\hfill
\vspace{-0.1in}
\end{figure*}

Since the QoE perceived by a client depends on the state of the buffer and the stalls experienced during video playback, we also study these individual components for the different policies under the different channel conditions. We can observe a similar trend when we compare the CDFs of the individual values for both these components in Figures \ref{fig:buf_comp} to \ref{fig:stall_cdf_bad}. We can clearly infer from the figures that the System-wide and the Auction-based policies ensure higher number of buffered seconds and lower duration of stalls under both channel conditions when compared to the other policies, even though the gap in performance is much less in the $Bad$ channel case.

We also compared the bid distributions of the clients in the Auction-based policy for two different client configurations. The first configuration had 6 clients whereas the second one had 3, with the total bandwidth allocation kept the same. The comparison of the two distributions is shown in Figure~\ref{fig:bid_dist}.  When there are more clients participating, resources are scarce and valuable, and so, clients tend to bid higher in order to get into the high priority queue and experience better QoE.  When the total number of clients is low, everyone experiences good QoE irrespective of the queue they are assigned to, and there is no incentive to bid higher.

\subsection{Dynamic Number of  Clients and Varying Channel}
In a realistic setting, the number of participating clients as well as the channel conditions do not remain static. Hence, we would like to evaluate the performance of the policies with dynamic number of clients under varying channel conditions. We vary the number of active clients in the system between 3 and 6 for the next set of experiments, under the assumption that a particular configuration of clients remains unchanged (in the same channel condition) for a duration of 30 minutes. Since this duration is large as compared to the policy decision period of 10 seconds, the use of a composite controller as described earlier is only slightly sub-optimal.

We fix a sequence of client configurations (number of active clients) under each channel condition for the evaluation of all policies. The first configuration consists of 6 clients under $Good$ channel conditions and 3 under $Bad$ channel conditions. We decrease the number of clients in the $Good$ scenario by 1 and increase that in the $Bad$ scenario by 1 for the next three intervals. The evolution of the average QoE for each of the policies for the above sequence is shown in Figure \ref{fig:qoe_evol}. The System-wide and Auction-based policies exhibit a high average QoE in most of the configurations except for the last where it is not possible to achieve a high QoE for the 6 clients in the $Bad$ channel. Even in such a scenario, the drop in QoE is more severe for the other policies. 

Decrease in the number of active clients under a constant bandwidth allocation results in the relaxation of the constraints for each individual client and hence, Round Robin and even the Vanilla approach perform better in some configurations. This improvement in performance, when compared with the constrained setting, can be seen in Figures \ref{fig:qoe_channel_cdf}, where the CDF curves of both policies are closer to those of the System-wide and the Auction-based policies. 

\section{Conclusion}
\label{section:conclusion}

In this paper, we described FlowBazaar, a platform for adaptive prioritization of flows in response to their self-declared values.  We showed that using an auction framework is able to a elicit truthful proxy for state in terms of the bid made for prioritized service.  Furthermore, the model needed at clients to make optimal bids is simply  the marginal transition kernel of the system, which is learned quite easily.  Using YouTube video streaming as an example application, we showed how  FlowBazaar is able to make the correct choices on which clients to prioritize, and actually ensured higher overall QoE than system-wide optimization (using a coarsely quantized state space). We also discovered ordering of state values that can be applied directly as a simple index policy (assuming truthful reporting of state).   Our future goal is to analytically characterize the nature of this index policy.

\bibliographystyle{IEEEtran}
\bibliography{references}

\begin{thebibliography}{10}
\providecommand{\url}[1]{#1}
\csname url@samestyle\endcsname
\providecommand{\newblock}{\relax}
\providecommand{\bibinfo}[2]{#2}
\providecommand{\BIBentrySTDinterwordspacing}{\spaceskip=0pt\relax}
\providecommand{\BIBentryALTinterwordstretchfactor}{4}
\providecommand{\BIBentryALTinterwordspacing}{\spaceskip=\fontdimen2\font plus
\BIBentryALTinterwordstretchfactor\fontdimen3\font minus
  \fontdimen4\font\relax}
\providecommand{\BIBforeignlanguage}[2]{{%
\expandafter\ifx\csname l@#1\endcsname\relax
\typeout{** WARNING: IEEEtran.bst: No hyphenation pattern has been}%
\typeout{** loaded for the language `#1'. Using the pattern for}%
\typeout{** the default language instead.}%
\else
\language=\csname l@#1\endcsname
\fi
#2}}
\providecommand{\BIBdecl}{\relax}
\BIBdecl

\bibitem{bhattacharyya2019qflow}
R.~Bhattacharyya, A.~Bura, D.~Rengarajan, M.~Rumuly, S.~Shakkottai,
  D.~Kalathil, R.~K. Mok, and A.~Dhamdhere, ``Qflow: A reinforcement learning
  approach to high {QoE} video streaming over wireless networks,'' \emph{arXiv
  preprint arXiv:1901.00959}, 2019.

\bibitem{7025402}
H.~Yeganeh, R.~Kordasiewicz, M.~Gallant, D.~Ghadiyaram, and A.~C. Bovik,
  ``Delivery quality score model for {I}nternet video,'' in \emph{Proceedings
  of IEEE ICIP}, 2014.

\bibitem{ManRam14}
M.~Manjrekar, V.~Ramaswamy, and S.~Shakkottai, ``A mean field game approach to
  scheduling in cellular systems,'' in \emph{Proceedings of IEEE INFOCOM},
  2014, pp. 1554--1562.

\bibitem{TasEph_92}
L.~Tassiulas and A.Ephermides, ``Stability properties of constrained queueing
  systems and scheduling policies for maximum throughput in multihop radio
  networks,'' \emph{IEEE Trans. Automat. Contr.}, vol.~37, no.~12, pp.
  1936--1948, 1992.

\bibitem{ErySriPer_05}
A.~Eryilmaz, R.~Srikant, and J.~Perkins, ``Stable scheduling policies for
  fading wireless channels,'' \emph{IEEE/ACM Trans. Network.}, vol.~13, pp.
  411--424, April 2005.

\bibitem{HouBor09}
I.~Hou, V.~Borkar, and P.~Kumar, ``{A theory of {QoS} for wireless},'' in
  \emph{Proceedings of IEEE INFOCOM}, 2009.

\bibitem{Auction06}
J.~Sun, E.~Modiano, and L.~Zheng, ``Wireless channel allocation using an
  auction algorithm,'' \emph{IEEE Journal on Selected Areas in Communications},
  vol.~24, no.~5, pp. 1085--1096, 2006.

\bibitem{HaSen12}
S.~Ha, S.~Sen, C.~Joe-Wong, Y.~Im, and M.~Chiang, ``T{UBE}: {T}ime-dependent
  pricing for mobile data,'' in \emph{Proceedings of ACM SIGCOMM}, 2012, pp.
  247--258.

\bibitem{CrossFlow1}
P.~Shome, M.~Yan, S.~M. Najafabad, N.~Mastronarde, and A.~Sprintson,
  ``Crossflow: A cross-layer architecture for {SDR} using {SDN} principles,''
  in \emph{Proceedings of IEEE NFV-SDN}, 2015.

\bibitem{CrossFlow2}
P.~Shome, J.~Modares, N.~Mastronarde, and A.~Sprintson, ``Enabling dynamic
  reconfigurability of {SDR}s using {SDN} principles,'' in \emph{Proceedings of
  Ad Hoc Networks}, 2017.

\bibitem{Muxi}
M.~Yan, J.~Casey, P.~Shome, A.~Sprintson, and A.~Sutton, ``{\AE}therflow:
  Principled wireless support in {SDN},'' in \emph{Proceedings of IEEE ICNP},
  2015.

\bibitem{aeroflux}
J.~Schulz-Zander, N.~Sarrar, and S.~Schmid, ``{AeroFlux}: {A} near-sighted
  controller architecture for software-defined wireless networks,'' in
  \emph{Proceedings of USENIX ONS}, 2014.

\bibitem{schulz2015opensdwn}
J.~Schulz-Zander, C.~Mayer, B.~Ciobotaru, S.~Schmid, and A.~Feldmann,
  ``Open{SDWN}: {P}rogrammatic control over home and enterprise {WiFi},'' in
  \emph{Proceedings of ACM SOSR}, 2015.

\bibitem{ericsson-mobility-report}
Ericsson, ``{Ericsson Mobility Report: On the Pulse of the Networked
  Society},''
  \url{https://www.ericsson.com/assets/local/mobility-report/documents/2015/ericsson-mobility-report-june-2015.pdf},
  2015.

\end{thebibliography}
\end{document}